\documentstyle[prl,aps]{revtex}
\begin{document}

\title{QUARK FRAGMENTATION FUNCTIONS
IN LOW-ENERGY CHIRAL THEORY\thanks
{This work is supported in part by funds provided by the U.S.
Department of Energy (D.O.E.) under contract \#DE-AC02-76ER03069
(MIT) and the National Science Foundation (Drexel).}}

\author{Xiangdong Ji$^{1}$ and Zheng-Kun Zhu$^{1,2}$}

\address{$^{1}$Center for Theoretical Physics,
Laboratory for Nuclear Science, and Department of Physics \\
Massachusetts Institute of Technology,
Cambridge, Massachusetts 02139--4307 \\
$^{2}$Department of Physics and Atmospheric Science\\
Drexel University, Philadelphia, PA 19104\\
{~}}

\date{MIT-CTP-2259. ~~~ Submitted to: {\it Phys. Rev. Lett.} ~~~ November 1993}

\maketitle

\thispagestyle{empty}

\begin{abstract}
We examine the physics content of fragmentation
functions for inclusive hadron production in
a quark jet and argue that it can be calculated in
low energy effective theories. As an example, we
present a calculation of $u$-quark fragmentation to
$\pi^+$ and $\pi^-$ mesons in the lowest order in
the chiral quark model. The comparison between
our result and experimental data is encouraging.
\end{abstract}


\widetext
Despite lack of a rigorous proof, many believe
that the color charges in Quantum Chromodynamics
(QCD) are permanently confined. The building
blocks of QCD, quarks and gluons, cannot emerge as
asymptotic states of the theory and thus are not
directly detectable in an experiment. Rather,
traces of energetic quarks and gluons in a hard
collision manifest in jets of hadrons
with highly correlated momenta.
Since their discovery in 1975, jets have
become bread-and-butter physics in
high-energy colliders.

Parton fragmentation refers to the
process of converting
high-energy, colored quarks and
gluons out of a hard scattering
into hadron jets observed in detectors.
Undoubtedly, this process
involves QCD physics at many different
scales and is rather complicated.
However, important developments
in perturbative QCD occurred in
the beginning of 80's, coupling
with rich experimental data taken from
high-energy colliders,
have taught us a great deal about what
is going on in fragmentation process \cite{MUE1}.
In a modern view, parton fragmentation
involves three key concepts:
separation of short and long
distance physics (factorization theorem or assumption),
perturbative evolution of partons
from high to low virtualities
(parton shower), and
non-perturbative fragmentation of partons
with virtuality of order of 1 GeV to hadrons
(hadronization). While the first two subjects
can be treated systematically in perturbation
theory, the last one is intrinsically non-perturbative
and is difficult to study directly in QCD.
In the past, phenomenological models, such as
Feynman-Field model \cite{FF} or Lund string model \cite{AND},
have been used to describe hadronization in Monte
Carlo simulations. Except for
heavy quarks \cite{XXX}, little
progress has been made on understanding
fragmentation physics
from the fundamental theory.

In this Letter we attempt to study hadronization
from a low energy effective theory,
focusing on calculating fragmentation functions
for inclusive hadron production.
In order for the reader to understand
the context of our calculation and to disperse
possible doubts over its relevance, we
begin with inclusive hadron production in $e^+e^-$
annihilation, for which a factorization
theorem can be proved rigorously
in perturbative QCD \cite{MUE2,CS1}. The theorem
asserts that in the leading order
in hard momentum the inclusive hadron is
produced by fragmentation of a
{\it single} quark without influence of others
(independent jet fragmentation). Consequently,
the fragmentation functions, which describe
hadron distributions in the jet,
can be expressed in terms of the matrix
elements of the quark field operator alone.
If similar factorization theorems can
be proved for other processes, the
same functions appear in the relevant
hadron-production cross sections.

Like parton distribution functions,
the parton fragmentation functions are
scale dependent, and the scale evolution is
governed by renormalization group equations\cite{MUE2}.
At low-energy scales, the fragmentation functions
contain no large momenta and shall be
calculable in low-energy models.
To illustrate this, we consider pion
production in a quark jet using
the chiral quark model of Manohar and
Georgi \cite{GM}. The tree level result for $\pi^+$ and
$\pi^-$ productions in a $u$-quark jet
shows an impressive similarity with
the EMC data when evolved to appropriate
energy scales. The higher-order corrections can
be taken into account systematically in
a chiral expansion.

To begin our discussion,
we consider the hadron tensor for
inclusive hadron production
in $e^+e^-$ annihilation,
\begin{equation}
    \hat  W_{\mu\nu} = {1\over 4\pi}\sum_X \int d^4\xi e^{iq\cdot \xi}
            \langle 0|J_\mu(\xi)|H(P)X\rangle
          \langle H(P)X|J_\nu(0)|0\rangle  $$
\label{W1}
\end{equation}
where $q$ is the moment of a time-like virtual photon,
$P$ is the moment of the observed hadron $H$ and
$X$ represents other unobserved hadrons and
is summed over. In the following
discussion, we choose a special coordinate system
defined by two light-cone vectors $p={\cal P}(1,0,0,1)$
and $n=1/(2{\cal P})(1,0,0,-1)$ with $p\cdot n=1$,
in which the hadron and photon momenta are collinear:
$P=p + nM^2/2 $, $q=p/z + \nu n$.
In the deep-inelastic limit ($Q^2 = q^2 \rightarrow \infty$,
$\nu=P\cdot q \rightarrow \infty$, and ${2\nu /Q^2 }
= z = $ finite), the factorization theorem
guarantees that the leading
contribution to the hadron tensor, neglecting the
calculable perturbative corrections,
comes from the diagrams in Fig.~1\cite{MUE2},
\begin{eqnarray}
     \hat W_{\mu\nu} & = &
       3\sum_a e_a^2 (\hat f_1^a(z,Q^2)
    +\hat f_1^{\bar a}(z, Q^2))/z^2  \nonumber\\
        &\times&
     \left[z(-g^{\mu\nu} + q^\mu q^\nu/q^2) -
            2/\nu(P^\mu-{z\over 2}q^\mu)
             (P^\nu-{z\over 2}q^\nu)\right]
\label{W2}
\end{eqnarray}
where $a$ sums over quark flavors and
\begin{equation}
     \hat f_1(z, \mu^2)
           = {1\over 4}z\int {d\lambda \over 2\pi}
           e^{-i\lambda/z} \langle 0 |
{\mathrel{\mathop{n\!\!\!/}}}_{\alpha\beta}\psi_\beta(0)|H(P)X\rangle
          \langle H(P)X|\bar \psi_\alpha(\lambda n)|0\rangle
\label{hf1}
\end{equation}
is the quark fragmentation function represented by
a quark-hadron four-point vertex in Fig.~1.
In Eq. (\ref{hf1}), $\mu^2$ labels the
renormalization-point dependence
and the light-cone gauge $A\cdot n =0$ has been
used (otherwise a gauge link has to be explicitly included to
ensure gauge invariance).

Except for a scale dependence, Eq. (\ref{W2}) is
the naive parton-model result proposed by Feynman
before QCD \cite{F}. It resembles a similar
prediction for the hadron structure functions
in deep-inelastic scattering, which can be justified
by the operator production expansion in QCD.
However, validity of Eq. (\ref{W2})
in QCD is somewhat more remarkable, for the color charges
are knowned to be confined at a scale of order 1 fm,
at which something must happen to ensure
the quark and antiquark keeping flying apart.
The factorization theorem says whatever mechanism
it is, it does not affect the hadron content of
a jet.

To understand better about this independent parton
fragmentation picture, we recall
the way that the gluon exchanges between
the quark and antiquark jets are treated
when the factorization theorem is proved\cite{CSS}.
There are two types of gluon exchanges which
are important in the so-called leading diagrams.
The first is the collinear gluons
emitted by a quark, with their momenta
parallel to the other quark. These gluons
are longitudinally polarized, and are
summed to a gauge link to make Eq. (3)
gauge invariant. The second is the soft gluons
which either are emitted by the jets or
link the two at large separations. With use of
the soft-gluon approximation and the Slavnov-Taylor
identities they can be factorized, and
are subsequently cancelled by unitarity
when final states, excluding the observed hadron,
are summed.

Thus, it appears that the study of inclusive hadron
production in $e^+e^-$ annihilation
reduces to evaluating Eq. (\ref{hf1}).
Notice the close similarity of
$\hat f_1(z)$ with the quark distribution function
$f_1(x)$ in a hadron of momentum $P$,
\begin{equation}
     f_1(x, \mu^2) = {1\over 2}\int {d\lambda \over 2\pi}
           e^{i\lambda x} \langle P|\bar \psi(0)
           {\mathrel{\mathop{n\!\!\!/}}}
           \psi(\lambda n)|P \rangle ~~.
\label{f1}
\end{equation}
Our experiences in calculating the latter provide
us with valuable insights in
calculating the former:
First, the fact that the spectators $X$ in Eq. (3) are
colored states is not a problem in a real
calculation. Similar colored intermediate
states occur in the distribution functions
if a complete set of states is inserted in-between
the quark fields. [In the MIT bag model, these
are di-quark states.] Second, the fragmentation
functions at $\mu^2$ less then 1 GeV$^2$ involve
only low-energy scales and are entirely
dominated by non-perturbative QCD physics. As
such, techniques useful for calculating the
parton distributions can in principle be used
to calculate the fragmentation functions.

The explicit sum over the spectators
can not be eliminated in the fragmentation
functions, even if one is only interested in
their moments. This renders lattice QCD and QCD
sum rule methods largely useless.
However, the low-energy chiral theory is
an exception. One version of the
theory particularly useful here
is the chiral quark model of
Manohar and Georgi \cite{GM}, which is an
effective theory of QCD at
scales between $\Lambda_\chi =4\pi f_\pi$,
the chiral symmetry breaking scale, and
$\Lambda_{\rm QCD}$, the QCD confinement scale.
Emergence of such a
theory at low energy can be argued as follows:
As an energy scale decreases below
$\Lambda_\chi$, the instability of the
perturbative QCD vacuum leads to
spontaneous breaking of the flavor
$SU(3)_L\times SU(3)_R$ chiral symmetry, creating
an octet of Goldstone bosons.
Meanwhile, the quarks and gluons acquires
their constituent masses through non-zero
vacuum condensates. The interactions between
the constituent quarks and gluons and Goldstone bosons
are determined by chiral dynamics and are
controlled by expansion of small parameters
$m_\pi^2/\Lambda_\chi^2$ and $k^2/\Lambda_\chi^2$,
where $k$ is a small momentum.

Matching the QCD quarks above
$\mu=\Lambda_\chi$ and the constituent
quarks below deserves
some explanations. First of all, to find
the exact matching conditions one has to
solve both QCD and the effective theory
around $\Lambda_\chi$ completely. Second, an
effective theory is effective only if matching
conditions are simple. In this study, we take
the most naive assumption that a QCD quark
{\it is} just a constituent quark
at the matching scale. This
is motivated by successes of similar assumptions
used in other constituent quark models.
We also note that the matching conditions
should be used in conjunction
with the way that the effective theory is treated.
We will return to this point later when we choose
a cut-off for ultra-violet momenta.

For simplicity, we will neglect
the gluon fragmentation at low energy scale,
because in the effective theory gluons interact
weakly with quarks, $\alpha_s^{\rm eff}\sim 0.3$.
This, of course, means that our result is unreliable
for small $z$, where hadrons are mostly produced
by bremsstrahlung gluons. In particular,
the so-called hump-back plateau in hadron spectra,
caused by intrajet coherence effects, is beyond
our scope \cite{MSDK}. Thus to the leading order,
the effective lagrangian for
quarks and Goldstone bosons is
\begin{equation}
      {\cal L} = \bar \psi (i
{\mathrel{\mathop{D\!\!\!/}}}
 +
{\mathrel{\mathop{V\!\!\!/}}} -m)
\psi + g_a\bar \psi
{\mathrel{\mathop{A\!\!\!/}}}
\gamma_5 \psi
\label{lagr}
\end{equation}
where $\psi$ carries implicit color, favor, and spin
indices. The vector and axial-vector fields are defined as,
\begin{equation}
          (V_{\mu}, A_{\mu}) = {i\over 2}(\xi^\dagger\partial_\mu
             \xi \pm \xi\partial_{\mu} \xi^{\dagger})
\end{equation}
where $\xi = \exp(i\pi/f_\pi)$ and $\pi = \sum_a \pi^a T^a$
with $f_\pi=93$ MeV and Tr$T^aT^b=\delta^{ab}/2$.
Under the chiral transformation:
\begin{eqnarray}
     \Sigma (= \xi^2) &\rightarrow& L\Sigma R^\dagger, \nonumber \\
     \xi  & \rightarrow & L\xi U^\dagger=U\xi R^\dagger, \nonumber \\
     \psi & \rightarrow & U\psi,
\end{eqnarray}
where $L$ and $R$ are group elements of $SU(3)_L\times SU(3)_R$,
${\cal L}$ is invariant.

Let us first consider the $\pi^+$ production
from a $u$-quark jet.
The momentum-space Feynman rules for
$f_1(z)$ can be derived
easily when re-writing $\hat f_1$ as,
\begin{equation}
     \hat f_1(z, \mu^2)
           = {1\over 4}z\int {dk^-d^2k_\perp
          \over (2\pi)^4} \int d^4\xi e^{-i\xi\cdot k}
           \sum_X \langle 0|
           \gamma^+_{\alpha\beta}
           \psi_\beta(0)|H(P)X\rangle
          \langle H(P)X|\bar \psi_\alpha(\xi)|0\rangle
\label{hf11}
\end{equation}
with $zk^+=p^+$, and each matrix element is
transformed to the interaction picture \cite{COLLINS}.
The lowest order diagram is shown in Fig.~2.
A simple calculation yields,
\begin{equation}
      \hat f_1^{\pi^+}(z) = {1\over 2z} g_a^2 \int
             {d k_\perp^2 \over (4\pi f_\pi)^2}
\end{equation}
In contrast to logarithmic theories, e.g., QED and QCD,
the pion transverse momentum integration
has no collinear singularity.
In large momentum region, it diverges quadratically.
In our calculation,
we cut off this type of integrations at
the scale $\Lambda_{\chi}$, beyond which
the effective theory ceases to be valid.
Of course, the result depends sensitively
on ways that the cut is imposed, more so than in
logarithmic theories. However, we believe that the
arbitrariness is cancelled when a choice is used in
conjunction with the corresponding
matching conditions. Here, we make
a simplest choice, $k_\perp^{\rm max}=\Lambda_\chi$. Thus,
\begin{equation}
      f_1^{\pi^+}(z, \Lambda_{\chi}^2) = {1\over 2z}g_a^2
\end{equation}
where $\Lambda_\chi =4\pi f_\pi$ has been used.
To confront this with experimental data, we
must evolve this to appropriate scales
using the Altarelli-Parisi equation \cite{FIELD}.
In Fig.~3, we show a comparison between the evolved
result ($g_a=0.75$) and the data from EMC measurement \cite{EMC}.
Considering the simplicity of the approach, we think
the agreement is impressive.

A more intricate case is $\pi^-$ production
from the $u$-quark jet, which is an unfavored
process. In the lowest
order, $\pi^-$ has to be produced together with a $\pi^+$
meson. There are two way to accomplish this:
The first is a sequential emission of
pions through the axial-vector coupling, and the
second is a sea-gull emission
through the vector coupling.
The two processes
interfere as shown in Fig.~4. The sign of
the interference term is completely determined by the
sign of the vector coupling, which in turn is
fixed by chiral symmetry.

The resulting expression for $\hat f_1(z)$ is
complicated and we evaluate it numerically.
A few salient features can be said briefly.
First, there is a $1/z$ divergence for the
longitudinal momentum integration of
$\pi^+$. A natural cut-off for
this is $m_\pi/{\Lambda_\chi} \sim 0.1 $,
the mass of pion over the scale of
the virtuality of the quark.
This is because pions cannot be
produced with a $z$ smaller than this
due to energy conservation. Second, the
numerical result shows a strong cancellation
between diagrams in Fig.~4a and~4b and the
interference diagrams
in Fig.~4c and~4d.
The cancelation is maximum if
$g_a=1/\sqrt{2}$, i.e., the vector
coupling is the square of the
axial-vector coupling.
In Fig.~5, we have shown the EMC data
and our relsult for $g_a=1.0$.
The fact that a slightly larger $g_a$ is needed to
reproduce the experimental data reflects the
oversimplified matching conditions we use.

Finally, we present a result for
the chiral-odd fragmentation function
$\hat e(z)$, for which there is no data available.
In Ref. \cite{JAFFEJI},
Jaffe and Ji pointed out the importance of
this fragmentation function in measuring
the transversity distribution of the nucleon
in deep-inelastic scattering. The QCD
definition for $\hat e(z)$ is ,
\begin{equation}
     \hat e_1(z, \mu^2)
           = {1\over 4M}z\int {d\lambda\over 2\pi}
            e^{-i\lambda/z}
            \langle 0|
           \psi_\alpha(0)|H(P)X\rangle
          \langle H(P)X|\bar \psi_\alpha(\xi)|0\rangle
\end{equation}
where $M$ is taken to be the nucleon mass.
A simple calculation in the chiral quark
model yields,
\begin{equation}
        \hat  e(z) = z\hat f_1(z) {m_q\over M}
        \sim {1\over 3}   z\hat f_1(z)
\end{equation}
where $m_q$ is the constituent quark mass.
Note that this relation is only true at the scale
$\Lambda_\chi$, beyond which $\hat e(z)$
evolves in a much complicated way (twist-three)
\cite{JI}.
However, as a rough estimate for $\hat e(z)$,
one can take (12) to be true beyond
the model, using it in conjunction
with the experimental data for $\hat f_1(z)$.

To summaries, we argue that the fragmentation
functions can be calculated
in low-energy effective theories.
As an example, we show how the pion fragmentation functions
are calculated in the chiral quark model.
The results seem to be encouraging. A study for
other fragmentation functions, including $K^\pm$
and extending possibly to $P(\bar P)$ productions,
will be presented elsewhere.

We thank J. Collion, M. Strikman for discussions
on factorization theorems and A. Manohar for discussions on
the chiral quark model. Z. Zhu would like to thank Professor
Da Hsuan Feng for his numerous encouragement and support.

\begin{figure}
\label{fig1}
\caption{The leading diagrams for inclusive
hadron production in $e^+e^-$ annihilation.}
\label{fig2}
\caption{The lowest-order diagram for $\pi^+$ production
in a $u$-quark jet.}
\label{fig3}
\caption{Fragmentaion function for $\pi^+$ production
in a $u$-quark jet. The data are taken from \protect\cite{EMC}.}
\label{fig4}
\caption{Same as Fig.~2, for $\pi^-$ production.}
\label{fig5}
\caption{Same as Fig.~3, for $\pi^-$ production.}
\end{figure}


\begin{references}
\bibitem{MUE1}
A. H. Mueller, {\it Perturbative Quantum Chromodynamics},
World Scientific, 1989.

\bibitem{FF}
R. D. Field and R. P. Feynman, Phys. Rev. D15 (1977) 2590.

\bibitem{AND}
B. Andersson, G. Gustafson, G. Ingelman, and T. Sjostrand,
Phys. Rep. 97 (1980) 31.

\bibitem{XXX}
C. Peterson et al., Phys. Rev. D27 (1983) 105;
E. Braaton, K. Cheung, and T. C. Yuan, Phys. Rev. D48 (1993)
5049; R. L. Jaffe and L. Randall, MIT CTP-preprint No. 2189.

\bibitem{MUE2}
A. H. Mueller, Physics Report 73 (1981) 237.

\bibitem{CS1}
J. Collins and D. Soper, Nucl. Phys. B185 (1981) 172.

\bibitem{GM}
A. Manohar and H. Georgi, Nucl. Phys. B234 (1984) 189.

\bibitem{F}
R. P. Feynman, {\it Photon-Hadron Interactions},
Benjamin-Cummings, 1972.

\bibitem{CSS}
J. C. Collins, D. E. Soper, and G. Sterman,
in {\it Perturbative Quantum Chromodynamics}, ed. by
A. H. Mueller, World Scientific, 1989.

\bibitem{MSDK}
A. H. Mueller, Phys. Lett. 104B (1981) 161;
Yu. L. Dokshitzer, V. S. Fadin, and V. A. Khoze,
Z. Phys. C15 (1982) 325.

\bibitem{COLLINS}
J. Collins, Nucl. Phys. B396 (1993) 161.

\bibitem{FIELD}
R. D. Field, {\it Applications of Perturbative QCD},
Addison-Wesley, 1989.

\bibitem{EMC}
M. Arneodo {\it et. al.} (the EMC collaboration),
Nucl. Phys. B321 (1989) 541; J. Aubert {\it et. al.}
(the EMC collaboration), Phys. Lett. B160 (1985) 417.

\bibitem{JAFFEJI}
R. L. Jaffe and X. Ji, Phys. Rev. Lett. 71 (1993) 2547.

\bibitem{JI}
X. Ji, MIT-CTP preprint No. 2219, to be appear in Phys. Rev. D, 1994.

\end{references}
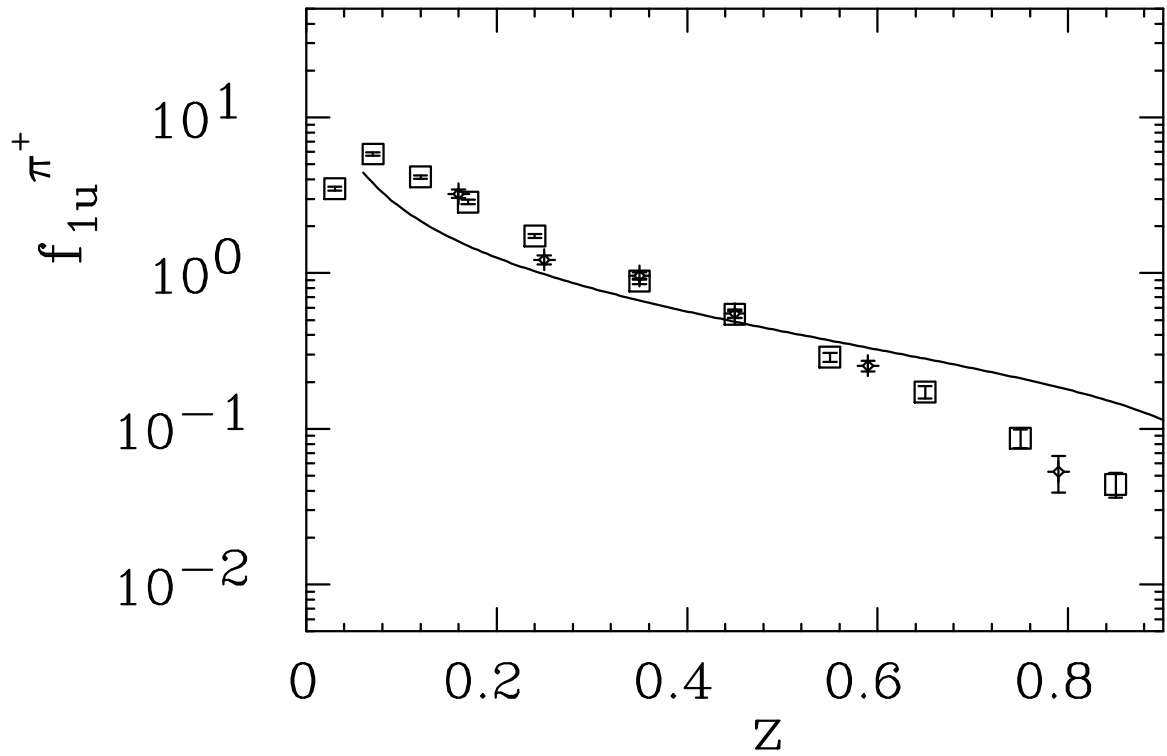
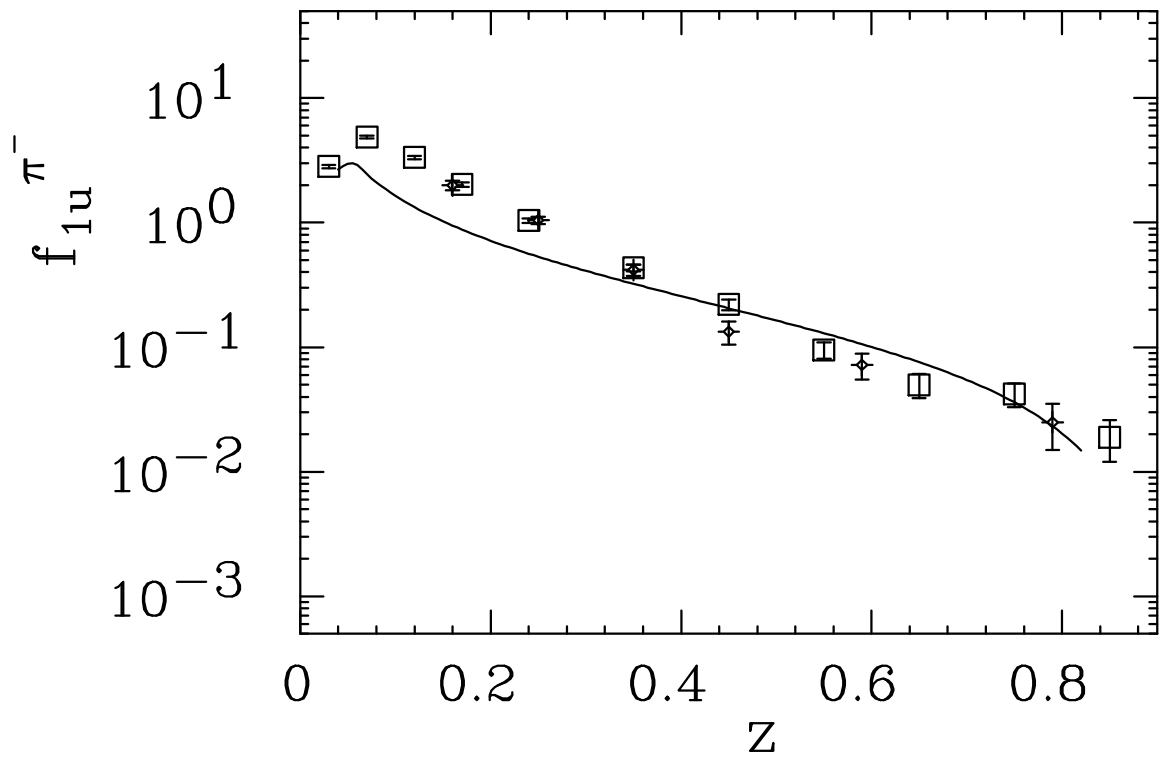
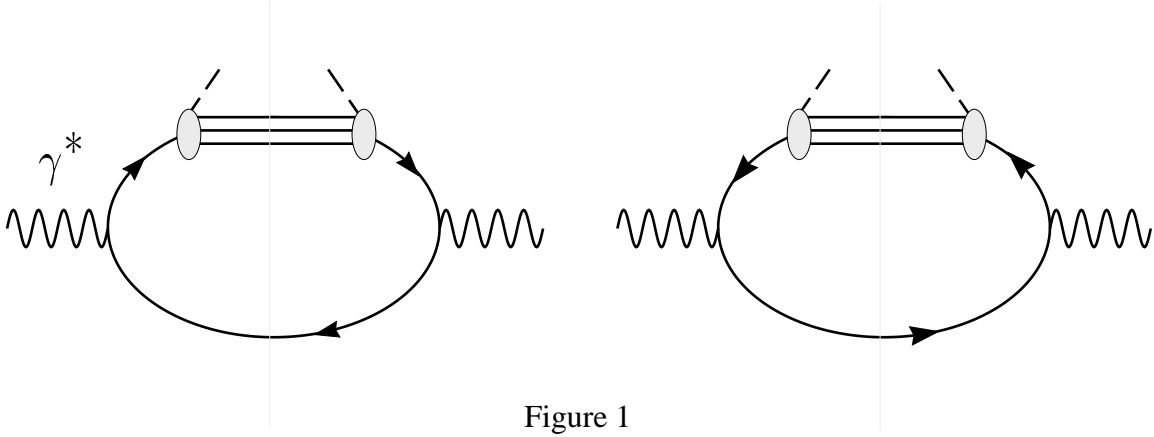
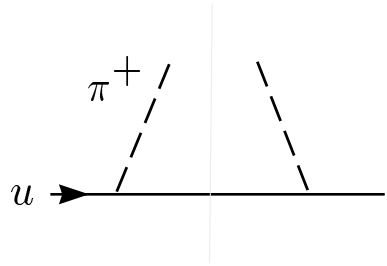
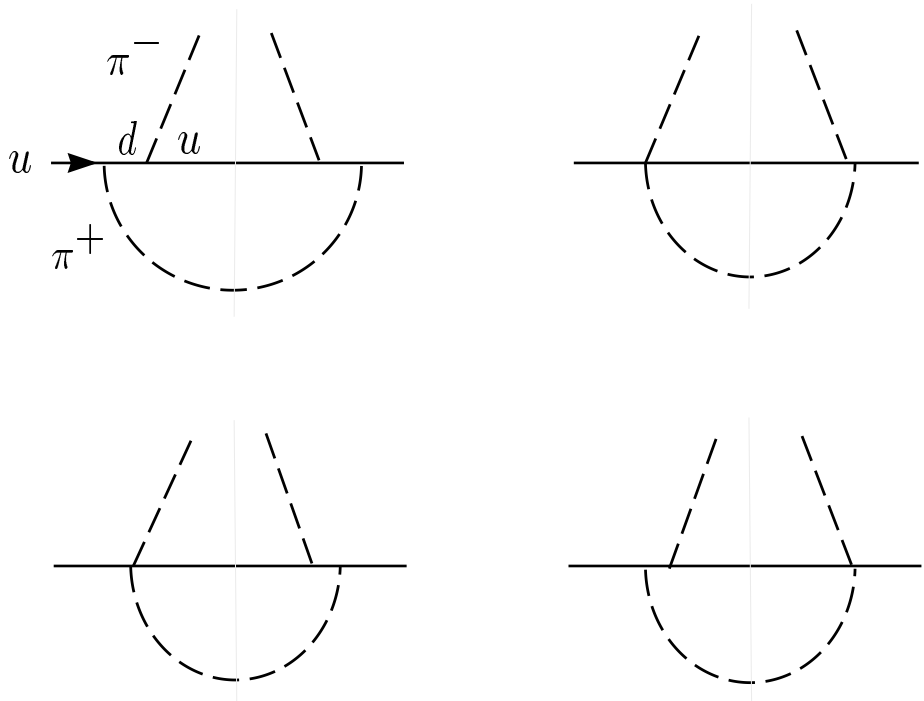
\end{document}